 \documentclass[final,5p,times,twocolumn]{elsarticle}

\usepackage{amssymb}
\usepackage{cite}
\usepackage{amsmath,amssymb,amsfonts}
\usepackage{algorithmic}
\usepackage{graphicx}
\usepackage{textcomp}
\usepackage{subfigure}
\usepackage{booktabs}
\usepackage{bbding}
\usepackage{makecell}
\usepackage{multirow}
\usepackage{cite}
\usepackage{array}
\usepackage{arydshln}
\usepackage{multirow}
\usepackage{bm}
\usepackage{color}
\usepackage{fancyhdr}
\begin{document}

\begin{frontmatter}

%Title, authors and addresses

%% use the tnoteref command within \title for footnotes;
%% use the tnotetext command for theassociated footnote;
%% use the fnref command within \author or \address for footnotes;
%% use the fntext command for theassociated footnote;
%% use the corref command within \author for corresponding author footnotes;
%% use the cortext command for theassociated footnote;
%% use the ead command for the email address,
%% and the form \ead[url] for the home page:
%% \title{Title\tnoteref{label1}}
%% \tnotetext[label1]{}
%% \author{Name\corref{cor1}\fnref{label2}}
%% \ead{email address}
%%% \ead[url]{home page}
%%% \fntext[label2]{}
%% \cortext[cor1]{}
%% \affiliation{organization={},
%%             addressline={},
%%             city={},
%%             postcode={},
%%             state={},
%%             country={}}
%% \fntext[label3]{}

%\pagestyle{plain}
\title{Convolutional dual graph Laplacian sparse coding}
\tnotetext[t1]{This work is supported by the National Natural Science Foundation of China (61771141, 62172098), and the Natural Science Foundation of Fujian Province (2020J01497,2021J01620).}
\cortext[cor1]{Corresponding author: Hang Cheng (e-mail: hcheng@fzu.edu.cn).}

%use optional labels to link authors explicitly to %addresses:
 \author[label1]{Xuefeng Peng}
 \affiliation[label1]{organization={School of Mathematics and Statistics},
             addressline={Fuzhou University},
             city={Fuzhou},
             postcode={Fujian 350108},
            country={China}}

 \affiliation[label2]{organization={College of Computer and Data Science},
             addressline={Fuzhou University},
             city={Fuzhou},
             postcode={Fujian 350108},
            country={China}}

\author[label2]{Fei Chen}
\author[label1]{Hang Cheng\corref{cor1}}
\author[label1]{Meiqing Wang}

%\affiliation{organization={},%Department and Organization
%            addressline={}, 
%            city={},
%            postcode={}, 
%            state={},
%            country={}}

\begin{abstract}
%% Text of abstract
In recent years, graph signal processing (GSP) technology has become popular in various fields, and graph Laplacian regularizers have also been introduced into convolutional sparse representation. This paper proposes a convolutional sparse representation model based on the dual graph Laplacian regularizer to ensure effective application of a dual graph signal smoothing prior on the rows and columns of input images.The graph Laplacian matrix contains the gradient information of the image and the similarity information between pixels, and can also describe the degree of change of the graph, so the image can be smoothed. Compared with the single graph smoothing prior, the dual graph has a simple structure, relaxes the conditions, and is more conducive to image restoration using the image signal prior. In this paper, this paper formulated the corresponding minimization problem using the proposed model, and subsequently used the alternating direction method of multiplication (ADMM) algorithm to solve it in the Fourier domain.Finally, using random Gaussian white noise for the denoising experiments. Compared with the single graph smoothing prior,the denoising results of the model with dual graph  smoothing prior proposed in this paper has fewer noise points and clearer texture.

\end{abstract}

%%Graphical abstract
%\begin{graphicalabstract}
%\includegraphics{grabs}
%\end{graphicalabstract}

%%Research highlights
%\begin{highlights}
%\item Research highlight 1
%\item Research highlight 2
%\end{highlights}

\begin{keyword}
convolutional sparse coding, image signal processing, image denoising, Laplacian
%% keywords here, in the form: keyword \sep keyword

%% PACS codes here, in the form: \PACS code \sep code

%% MSC codes here, in the form: \MSC code \sep code
%% or \MSC[2008] code \sep code (2000 is the default)

\end{keyword}

\end{frontmatter}

%% \linenumbers

%% main text
\section{Introduction}
\label{}
Distortion or missing data problems often occur in obtained signals and images owing to the limitations of image sensing, compression artifacts, and transmission errors [1]. This paper examines the problem associated with image restoration [2][3], that is, the reconstruction or estimation of missing data in images, which are immeasurable or corrupted by noise. Fundamentally, image restoration is an uncertain inverse problem-solving process, and appropriate signal priors are necessary to regulate this underdetermined inverse problem. The use of prior information about images can be model-based, which explicitly represents priors, such as total variation (TV) [4][5] and sparsity priors [6][7], or neural network-based, which explores priors by learning, such as deep learning (DL) [8]. Generally, DL-based methods provide superior performance compared to model-based methods in image restoration. However, pure DL schemes require large training data with annotations that are time-consuming and difficult to obtain. In contrast, model-based image restoration methods are relatively straightforward.

If the positions of the pixels in an image are considered as nodes and the neighborhood relations between pixel points as edges, then the geometry of the image can be considered as a graph, and the color or brightness information of the image can be considered as a signal defined on the graph. Thus, graph signal processing (GSP) technology [9][10][11] can be used for image restoration. The GSP technology performs the common signal operations, such as filtering, smoothing, and Fourier transform, on the signals defined on a graph. It then combines the signal information defined on the graph with the common graph operations, such as the graph Laplacian matrix [12], to form the prior information of the graph. Owing to these merits, GSP has become a state-of-the-art method for analyzing irregular structures [13][14]. 

Model-based image restoration methods generally include fidelity and regularization terms. Prior information is typically represented by a regularization term. A commonly used fidelity term is convolutional sparse representation [15][16]. Similar to the Fourier transform or wavelet transform, a signal can be decomposed and represented by a linear combination of a small number of dictionary elements. “Sparse” denotes that there are many dictionary elements present; however, only a few are necessary for a given signal. When the linear combination is substituted by a convolutional operation, the sparse representation is extended to convolutional sparse representation or convolutional sparse coding (CSC), which allows a sparsity prior to be applied globally rather than locally [17]. Common CSC models used for image restoration include the convolutional basis pursuit denoising (CBPDN) model [18] and online convolutional sparse coding [19].

Liu and Wohlberg first combined the CSC model with graph signal information to address the image inpainting problem [20]. In this method, an image was partitioned into overlapping patches, each of which corresponded to a vertex of the graph, and the similarity information between the image patches was represented by the edge weights of the graph. The graph signals defined on the vertices were the color or brightness vectors corresponding to the image patches. The Dirichlet energy of the graph signal was then used as a regularization term owing to its non-local smoothing properties.

Another method of constructing a graph signal through an image is the dual graph method [21][22]. The dual graph includes a row graph and a column graph. A vertex of the row graph corresponds to a row of the image, and a vertex of the column graph corresponds to a column of the image. The definitions of edge weights and signal information are similar to those of the single graph defined by [20]. The structure of the dual graph is simple compared to that of the single graph, and it relaxes the conditions, which is beneficial in image restoration when using graph signal priors.

In this paper, the dual graph signal prior was combined with the CBPDN model for image restoration. The contribution of this paper is twofold.
\begin{enumerate}
\item Introducing the dual graph prior information into the CSC model can solve the shortcomings of the $l^{1}$ regularizer in CSC and also compensate for the insufficiency of the single graph prior in [20]. To the best of our knowledge, this is the first paper to apply dual graph prior information to CSC.
\item A CSC model with prior information of the dual graph is proposed in this paper. The image denoising results of the proposed model are significantly better than those of the CSC model with single graph prior information, in terms of objective evaluation and visual quality.
\end{enumerate}

The outline of this paper is as follows. Section 2 introduces the related work and motivation of this article. Section 3 introduces the CSC model with a dual graph smoothing prior. Section 4 discusses the solution process of the model used in this peper. The experimental results and analysis are presented in Section 5. Finally, Section 6 provides a summary of this paper. 
\section{Related Work}
CSC is a classic image restoration model. It is a new variant of sparse coding, in which the entire signal or image is decomposed into a set of filters and the convolutional sum of the corresponding coefficient maps. The coefficient maps and signals or images are of the same size and have corresponding dictionary filters. A prominent formula for this problem is the CBPDN:
\begin{equation}\mathop {\min }\limits_{{{\mathbf{x}}_k}} \frac{1}{2}\left\| {\sum\limits_k {{{\mathbf{d}}_k}*{{\mathbf{x}}_k} - {\mathbf{s}}} } \right\|_2^2 + \lambda \sum\limits_k {{{\left\| {{{\mathbf{x}}_k}} \right\|}_1}} \label{eq}\end{equation}
where $\{ {{\mathbf{d}}_k}\} $ is a set of dictionary filters, $ * $ denotes convolution, and $\{ {{\mathbf{x}}_k}\} $ is a set of coefficient maps.

The $l^{1}$ regularizer in the standard CSC model can lead to high spatial sparsity of coefficient mapping, which consequently reduces the number of training image patches that play a role in the formation of the dictionary. Wohlberg et al. proposed an improved form of CBPDN that includes an additional regularization term based on the Laplacian of the non-local graph of the image. The graph is constructed by introducing each image patch corresponding to a vertex of the graph into the graph Laplacian regularizer $\sum\limits_k {\left\langle {{{\mathbf{x}}_k},{\mathbf{L}}{{\mathbf{x}}_k}} \right\rangle } $, as shown in Equation (2). Thus, the regularizer is an explicit penalty to enforce similar image patches with similar sparse representations.
\begin{equation}\mathop {\min }\limits_{{{\mathbf{x}}_k}} \frac{1}{2}\left\| {\sum\limits_k {{{\mathbf{d}}_k}*{{\mathbf{x}}_k} - {\mathbf{s}}} } \right\|_2^2 + \lambda \sum\limits_k {{{\left\| {{{\mathbf{x}}_k}} \right\|}_1}\rm{ + }}\frac{\mu }{2}\sum\limits_k {\left\langle {{{\mathbf{x}}_k},{\mathbf{L}}{{\mathbf{x}}_k}} \right\rangle} 
 \end{equation}

where ${\mathbf{L}}$ is a graph Laplacian matrix for a non-local graph.
However, this method has its shortcomings. In the image, the rows and columns share similar information. The above method ignores the adjacent information between the image rows and columns. Specifically, it uses the local information of the image, without fully considering the information and similarity between the rows and columns of the image. Therefore, in this paper, this article propose another variant of CBPDN that uses dual graph smoothing priors to regularize feature maps. The dual graph Laplacian regularizer [21] is introduced by using the rows and columns of the image as the vertices to form the row and column graphs, respectively. This variant effectively utilizes the adjacent information of image rows and columns. This paper adopted ideas from manifold learning [23] to constrain the solution for this article to be smooth on the graphs to implicitly enforce row and column proximities.
\section{Proposed Model}
In this section, this paper discuss the processes for constructing row  and column graphs, and introduce the CSC model with a dual graph smoothing prior.
\subsection{Column Graph}
This article uses $\mathbf{s} \in \mathbb{R}^{M \times N}$ to represent the original image matrix and consider each column of ${\mathbf{s}}$ as a node to form an undirected weighted column graph ${{\cal G}_c} = \{ {{\cal V}_c},{{\cal E}_c},{{\cal W}_c}\} $, its nodes ${{\cal V}_c} \in \{ 1,2, \cdots ,N\} $, edges ${{\cal E}_c} \in {{\cal V}^c} \times {{\cal V}^c}$, and weight matrix $\mathbf{W}_{c} \in \mathbb{R}^{N \times N}$. The weight matrix $\mathbf{W}_{c}$ describes the pairwise similarity between nodes in $\mathcal{G}_{c}$. $\mathbf{W}_{c}$ is defined as: 
\begin{equation}{{\mathbf{W}}_c}(i,j) = \left\{ {\begin{array}{*{20}{c}}
{\exp ( - \frac{{{{\rm{d}}_{ij}}^2}}{{{\delta ^2}}}),j \in {{\cal N}_i}}\\
{0,o.w.}
\end{array}} \right. \end{equation}
where $\mathcal{N}_{i}$ is the neighborhood defined by $i$ as the center pixel, $\delta$ is a factor that controls the shape of the function for weight calculation, and ${{\rm{d}}_{ij}}$  is a measure of the cosine distance [24] between vertex $i$ and vertex $j$, with a value range of [-1,1]. The more similar the two vertices, the closer they are to one. Thus, the graph $\mathcal{G}_{c}$ is completely and uniquely determined by the weight matrix $\mathbf{W}_{c}$.

Next, this article define the graph Laplacian matrix corresponding to column graph $\mathcal{G}_{c}$ as $\mathbf{L}_{c}=\mathbf{D}_{c}-\mathbf{W}_{c} \in \mathbb{R}^{N \times N}$, where the degree matrix $\mathbf{D}_{c}$ is a diagonal matrix with entries $
\mathbf{D}_{c}(i, i)=\sum_{j \neq i} \mathbf{W}_{\mathbf{c}}(i, i)
$. Given that the edge weights are non-negative in (3), it can be proved that the matrix $\mathbf{L}_{c}$ is positive semi-definite (PSD). This paper consider ${{\mathbf{x}}_k}$ as a collection of M-dimensional column vectors denoted by subscripts ${{\mathbf{x}}_k} = ({{\mathbf{x}}_{k,1}},{{\mathbf{x}}_{k,2}}, \cdots ,{{\mathbf{x}}_{k,N}})$. Regarding the columns ${{\mathbf{x}}_{k,1}},{{\mathbf{x}}_{k,2}}, \cdots ,{{\mathbf{x}}_{k,N}}$ as a vector-valued function defined on the vertices ${{\cal V}_c}$, the assumption of the smoothness of the implies that ${{\mathbf{x}}_{k,i}} \approx {{\mathbf{x}}_{k,j}}$, if $(i,j) \in {{\cal E}_c}$. Specifically, \begin{equation}\sum\limits_{i,j \in {{\cal V}_c}} {{{\mathbf{W}}_c}} (i,j){({{\mathbf{x}}_{k,i}} - {{\mathbf{x}}_{k,j}})^2} = {\mathop{\rm tr}\nolimits} ({{\mathbf{x}}_k}{{\mathbf{L}}_c}{\mathbf{x}}_k^{\rm T}) \end{equation}
needs to be small. Equation (4) is referred to as the column graph Laplacian regularizer (CGLR) of the feature maps ${{\mathbf{x}}_k}$. When the nodes $i$ and $j$ are more similar, ${{\mathbf{W}}_c}(i,j)$ is smaller; thus, when similar vertices present similar values, Equation (4) becomes small.
\subsection{Row Graph}
Row graphs can be defined similarly to the column graphs. Each row of ${\mathbf{s}}$ is considered as a node to form an undirected weighted row graph ${{\cal G}_r} = \{ {{\cal V}_r},{{\cal E}_r},{{\cal W}_r}\} $, with nodes ${{\cal V}_r} = \{ 1,2, \cdots ,M\} $, edges ${{\cal E}_r} \in {{\cal V}^r} \times {{\cal V}^r}$, and weight matrix $\mathbf{W}_{r} \in \mathbb{R}^{M \times M}$. The graph Laplacian matrix of ${{\cal G}_r}$ is $\mathbf{L}_{r}=\mathbf{D}_{r}-\mathbf{W}_{r} \in \mathbb{R}^{M \times M}$. This paper consider ${{\mathbf{x}}_k}$ as a collection of N-dimensional row vectors denoted by subscripts ${{\mathbf{x}}_k} = {(\left(\mathbf{x}_{k}^{1}\right)^{\mathrm{T}},\left(\mathbf{x}_{k}^{2}\right)^{\mathrm{T}}, \cdots ,\left(\mathbf{x}_{k}^{M}\right))^{\mathrm{T}}}$. Regarding the rows $\left(\mathbf{x}_{k}^{1}\right)^{\mathrm{T}},\left(\mathbf{x}_{k}^{2}\right)^{\mathrm{T}}, \cdots,\left(\mathbf{x}_{k}^{M}\right)^{\mathrm{T}}$   
 as a vector-valued function defined on the vertices 
 ${{\cal V}_r}$, the row graph Laplacian regularizer (RGLR) of the feature maps ${{\mathbf{x}}_k}$ is ${\mathop{\rm tr}\nolimits}({\mathbf{x}}_k^{\rm T}{{\mathbf{L}}_r}{{\mathbf{x}}_k})$.
 \subsection{CSC Model with Dual Graph Smoothing Prior}
By combining CGLR and RGLR, the dual graph Laplacian regularizer (DGLR) of the feature maps ${{\mathbf{x}}_k}$ can be obtained.
\begin{equation}{\mathop{\rm G}\nolimits} ({{\mathbf{x}}_k}) = {\mathop{\rm tr}\nolimits} ({{\mathbf{x}}_k}{{\mathbf{L}}_c}{\mathbf{x}}_k^{\rm T}) + {\mathop{\rm tr}\nolimits} ({\mathbf{x}}_k^{\rm T}{{\mathbf{L}}_r}{{\mathbf{x}}_k})\end{equation}
The CSC model with DGLR is as follows:
\[\mathop {\min }\limits_{\{ {{\mathbf{x}}_k}\} } \frac{1}{2}\left\| {\sum\limits_k {{{\mathbf{d}}_k}*{{\mathbf{x}}_k} - {\mathbf{s}}} } \right\|_2^2 + \lambda \sum\limits_k {{{\left\| {{{\mathbf{x}}_k}} \right\|}_1}}  + \frac{\alpha }{2}\sum\limits_k {{\mathop{\rm tr}\nolimits} ({{\mathbf{x}}_k}{{\mathbf{L}}_c}{\mathbf{x}}_k^{\rm T})}  \]
\begin{equation}+ \frac{\beta }{2}\sum\limits_k {{\mathop{\rm tr}\nolimits} ({\mathbf{x}}_k^{\rm T}{{\mathbf{L}}_r}{{\mathbf{x}}_k})} \end{equation}
$\lambda$, $\alpha$, and $\beta $ are trade-off parameters between the three regularization terms.

\section{Energy Optimization}
This paper employed the alternating direction method of multiplication (ADMM) [25] framework to solve Equation (6). Based on the ADMM single split method, Equation (6) is rewritten by separating the variable $\{ {{\mathbf{y}}_k}\} $,
\[\mathop {\min }\limits_{\{ {{\mathbf{x}}_k}\} } \frac{1}{2}\left\| {\sum\limits_k {{{\mathbf{d}}_k}*{{\mathbf{x}}_k} - {\mathbf{s}}} } \right\|_2^2 + \lambda \sum\limits_k {{{\left\| {{{\mathbf{y}}_k}} \right\|}_1}}  + \frac{\alpha }{2}\sum\limits_k {{\mathop{\rm tr}\nolimits} ({{\mathbf{y}}_k}{{\mathbf{L}}_c}{\mathbf{y}}_k^{\rm T})} \]
\begin{equation} + \frac{\beta }{2}\sum\limits_k {{\mathop{\rm tr}\nolimits} ({\mathbf{y}}_k^{\rm T}{{\mathbf{L}}_r}{{\mathbf{y}}_k})} 	\qquad {\mathop{\rm s}\nolimits} .t.\;{{\mathbf{x}}_k} = {{\mathbf{y}}_k}\end{equation}
Subsequently, equation (7) can be decomposed into the following three sub-problems:

\noindent(a) First, we fix the auxiliary variable$\{ {{\mathbf{y}}_k}\} $ and the multiplier $\{ {{\mathbf{u}}_k}\} $, and update the coefficient maps $\{ {{\mathbf{x}}_k}\} $.
\[{\{ {{\mathbf{x}}_k}\} ^{(t + 1)}} = \mathop {\min }\limits_{\{ {{\mathbf{x}}_k}\} } \frac{1}{2}\left\| {\sum\limits_k {{{\mathbf{d}}_k}*{{\mathbf{x}}_k} - {\mathbf{s}}} } \right\|_{\rm{2}}^{\rm{2}}\]
\begin{equation} + \frac{\rho }{2}\sum\limits_k {\left\| {{{\mathbf{x}}_k} - {\mathbf{y}}_k^{(t)} + {\mathbf{u}}_k^{(t)}} \right\|_{\rm{2}}^{\rm{2}}} \end{equation}
(b) We then fix the coefficient maps $\left\{ {{{\mathbf{x}}_k}} \right\}$ and the multiplier $\left\{ {{{\mathbf{u}}_k}} \right\}$, and update the auxiliary variable $\left\{ {{{\mathbf{y}}_k}} \right\}$.
\[{\{ {{\mathbf{y}}_k}\} ^{(t + 1)}} = \mathop {\min }\limits_{\{ {{\mathbf{y}}_k}\} } \lambda \sum\limits_k {{{\left\| {{{\mathbf{y}}_k}} \right\|}_1}}  + \frac{\alpha }{2}\sum\limits_k {{\mathop{\rm tr}\nolimits} ({{\mathbf{y}}_k}{{\mathbf{L}}_c}{\mathbf{y}}_k^{\rm T}) } \]
\begin{equation}+\frac{\beta }{2}\sum\limits_k {{\mathop{\rm tr}\nolimits} ({\mathbf{y}}_k^{\rm T}{{\mathbf{L}}_r}{{\mathbf{y}}_k})}  + \frac{\rho }{2}\sum\limits_k {\left\| {{\mathbf{x}}_k^{(t + 1)} - {{\mathbf{y}}_k} + {\mathbf{u}}_k^{(t)}} \right\|_2^2} \end{equation}
(c) Finally, the fixed coefficient maps $\left\{ {{{\mathbf{x}}_k}} \right\}$, the auxiliary variable $\{ {{\mathbf{y}}_k}\} $, and the multiplier $\left\{ {{{\mathbf{u}}_k}} \right\}$ are updated.
\begin{equation}{{\rm{\{ }}{{\mathbf{u}}_k}\} ^{(t + 1)}} = {{\mathbf{u}}_k} + {\mathbf{x}}_k^{(t + 1)} - {\mathbf{y}}_k^{(t + 1)}\end{equation}
The update of $\left\{ {{{\mathbf{x}}_k}} \right\}$ is similar to that in the standard convolutional learning case, and it can be solved effectively in the Fourier domain [26]. $\{ {{\mathbf{y}}_k}\} $ can also be solved using the ADMM algorithm. We separate the variable $\{ {{\mathbf{z}}_k}\} $ and rewrite Equation (9).
\[{\{ {{\mathbf{y}}_k}\} ^{(t + 1)}} = \mathop {\min }\limits_{\{ {{\mathbf{y}}_k}\} } \lambda \sum\limits_k {{{\left\| {{{\mathbf{y}}_k}} \right\|}_1}}  + \frac{\alpha }{2}\sum\limits_k {{\mathop{\rm tr}\nolimits} ({{\mathbf{z}}_k}{{\mathbf{L}}_c}{\mathbf{z}}_k^{\rm T})  } \]
\[+\frac{\beta }{2}\sum\limits_k {{\mathop{\rm tr}\nolimits} ({\mathbf{z}}_k^{\rm T}{{\mathbf{L}}_r}{{\mathbf{z}}_k})}  + \frac{\rho }{2}\sum\limits_k {\left\| {{\mathbf{x}}_k^{(t + 1)} - {{\mathbf{y}}_k} + {\mathbf{u}}_k^{(t)}} \right\|_2^2} \]
\begin{equation}{\mathop{\rm s}\nolimits} .t.\;{{\mathbf{y}}_k} = {{\mathbf{z}}_k}\end{equation}
Subsequently, we introduce the Lagrange multiplier $\left\{ {{{\mathbf{v}}_k}} \right\}$, and Equation (11) can be decomposed into the following three sub-problems:
\[{\{ {{\mathbf{y}}_k}\} ^{(t + 1)}} = \mathop {\min }\limits_{\{ {{\mathbf{y}}_k}\} } \lambda \sum\limits_k {{{\left\| {{{\mathbf{y}}_k}} \right\|}_1}}  + \frac{\rho }{2}\sum\limits_k {\left\| {{\mathbf{x}}_k^{(t + 1)} - {{\mathbf{y}}_k} + {\mathbf{u}}_k^{(t)}} \right\|_2^2} \]
\begin{equation} + \frac{\eta }{2}\sum\limits_k {\left\| {{{\mathbf{y}}_k}-{\mathbf{  z}}_k^{(t)} + {\mathbf{v}}_k^{(t)}} \right\|_2^2} \end{equation}
\[{\{ {{\mathbf{z}}_k}\} ^{(t + 1)}} = \mathop {\min }\limits_{\{ {{\mathbf{z}}_k}\} } \frac{\alpha }{2}\sum\limits_k {{\mathop{\rm tr}\nolimits} ({{\mathbf{z}}_k}{{\mathbf{L}}_c}{\mathbf{z}}_k^{\rm T}) + } \frac{\beta }{2} \sum\limits_k {{\mathop{\rm tr}\nolimits}({\mathbf{z}}_k^{\rm T}{{\mathbf{L}}_r}{{\mathbf{z}}_k})} \]
\begin{equation} + \frac{\eta }{2}\sum\limits_k {\left\| {{\mathbf{y}}_k^{(t + 1)} - {{\mathbf{z}}_k} + {\mathbf{v}}_k^{(t)}} \right\|_2^2} \end{equation}
\begin{equation}{{\rm{\{ }}{{\mathbf{v}}_k}\} ^{(t + 1)}} = {{\mathbf{v}}_k} + {\mathbf{y}}_k^{(t + 1)} - {\mathbf{z}}_k^{(t + 1)}\end{equation}
$\left\{ {{{\mathbf{y}}_k}} \right\}$ can be efficiently solved using the soft-threshold method. The update of $\left\{ {{{\mathbf{z}}_k}} \right\}$ is a convex optimization problem; hence, a closed solution can be directly obtained. To solve the unconstrained QP problem (13), this article refer to [21] to derive the objective function for ${{\mathbf{z}}_k}$, then set it to zero, and solve for ${{\mathbf{z}}_k}$ to obtain the linear equation system of unknown ${\mathop{\rm vec}\nolimits} ({\mathbf{z}}_k^*)$:
\[(\alpha {{\mathbf{L}}_c} \otimes {{\mathbf{I}}_M} + \beta {{\mathbf{I}}_N} \otimes {{\mathbf{L}}_r} + \eta {\mathbf{I}}){\mathop{\rm vec}\nolimits} ({\mathbf{z}}_k^*)\]
\begin{equation} = \eta {\mathop{\rm vec}\nolimits} ({{\mathbf{y}}_k} + {{\mathbf{v}}_k}),k = 1,2, \cdots \end{equation}
Here, ${\mathop{\rm vec}\nolimits} ( \cdot )$ refers to the vector form of a matrix superimposed by columns, and $ \otimes $ is a defined operator. The calculation result of $\mathbf{Q}=\alpha \mathbf{L}_{c} \otimes \mathbf{I}_{M}+\beta \mathbf{I}_{N} \otimes \mathbf{L}_{r}, \mathbf{Q} \in \mathbb{R}^{M N \times M N}$ is represented by Equation (16). Because the coefficient matrix ${\mathbf{Q}}$ is generally symmetric, sparse, and positive definite, a plethora of mature numerical linear algebra methods such as conjugate gradient (CG) [27] can be used to solve equation (15) effectively. A unique advantage of using the dual graph signal smoothness prior in Equation (1) is that it requires only solving a system of linear equations when computing its solution. For the relative residual stopping rule, this paper refer to [28].
\[
\mathbf{Q}=\alpha\left[\begin{array}{cccc}
l_{11} \mathbf{I}_{M} & l_{12} \mathbf{I}_{M} & \cdots & l_{1 N} \mathbf{I}_{M} \\
l_{21} \mathbf{I}_{M} & l_{22} \mathbf{I}_{M} & \cdots & l_{2 N} \mathbf{I}_{M} \\
\vdots & \vdots & \ddots & \vdots \\
l_{N 1} \mathbf{I}_{M} & l_{N 2} \mathbf{I}_{M} & \cdots & l_{N N} \mathbf{I}_{M}
\end{array}\right]\]
\begin{equation}+\beta\left[\begin{array}{cccc}
\mathbf{L}_{r} & 0 & \cdots & 0 \\
0 & \mathbf{L}_{r} & \cdots & 0 \\
\vdots & \vdots & \ddots & \vdots \\
0 & 0 & \cdots & \mathbf{L}_{r}
\end{array}\right]
\end{equation}
Here, $l_{i j}, i, j=1,2, \cdots, N$ is the element in the row and column of the matrix ${{\mathbf{L}}_c}$.
\section{Experiment}
This paper used random Gaussian white noise for the denoising experiments to verify the performance of the proposed model in image restoration. In this paper, two types of datasets, namely, "kodim" dataset and "drone" image dataset, were used to perform Gaussian white noise denoising experiments. The experiments were run on a computer configured with IntelCorei7-6700U and a CPU of 3.41 GHz, using Matlab 2018b version.

\begin{figure}[htbp]
\setlength{\abovecaptionskip}{0cm}

 \begin{center}
\includegraphics[scale=0.4]{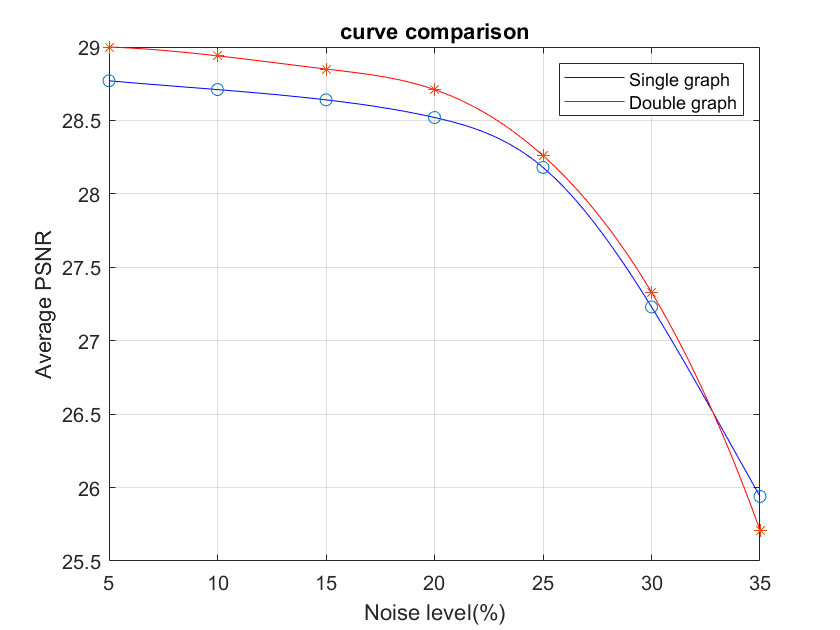}
\end{center}

\caption{The average denoising results of our model and SCSC under different noise levels.}

\label{fig1}

\end{figure}
 
 To evaluate the robustness of this model to different noise levels, noise levels of 5$\%$, 10$\%$, 15$\%$, 20$\%$, 25$\%$, 30$\%$, and 35$\%$, were used to distort the clean image. Separate denoising experiments were performed for each level of noise corruption. Furthermore, this article compared the denoising performance of this model with three other models: 1) CSC; 2) CSC based on a single graph (SCSC); and 3) three-dimensional block matching method (BM3D), which is a popular denoising technology [29]. The proposed CSC model based on the dual graph is denoted as DCSC herein. SCSC and DCSC are extensions of the CSC model, and were compared based on innovation. BM3D is the most popular denoising method among current methods. The addition of BM3D as a comparison method was based on its effective denoising effect.
This paper used the PSNR value of the image and visualization as a metric to compare the denoising performance of the models. 

Each model had several parameters to be selected during the verification phase of the experiment. Both SCSC and DCSC added different regular terms to the CSC model. To perform a fair comparison of the three models, this paper first adjusted the parameters of the CSC to the optimal values, and subsequently adjusted the remaining parameters in the SCSC and DCSC models. After debugging, this article set the parameters of (2) and (6) to $\lambda=0.28$, $\quad \mu=0.1$ and$\quad \alpha=\beta=0.2$. In the convolutional sparse coding model, the dictionary needs to be trained through the training set. In order to avoid the amount of calculation and to ensure the fairness of the experiment, this paper uses the same independently trained 12×12×36 dictionary for all models to solve the model.

\begin{table*}
\setlength{\abovecaptionskip}{-1.5cm}
\setlength{\belowcaptionskip}{0.1cm}
\renewcommand\arraystretch{1.2}
\caption{Comparison of PSNR values of denoising results of different methods under different noise levels}
\label{table1}
\centering
\begin{tabular}{|m{1.5cm}<{\centering}|m{1.5cm}<{\centering}|m{1.5cm}<{\centering}| m{1.5cm}<{\centering}|m{1.5cm}<{\centering}|m{1.5cm}<{\centering}|m{1.5cm}<{\centering}|m{1.5cm}<{\centering}|m{1.5cm}<{\centering}|}
  \hline
 \multirow{2}{*}{Image} &  \multicolumn{4} {|c|} { $\delta {\rm{ = }}15$ }&   \multicolumn{4} {|c|} { $\delta {\rm{ = }}20$}   \\    
  \cline{2-9}
           
                        & CSC   & SCSC  & BM3D  & Ours  & CSC   & SCSC  & BM3D  & Ours   \\
   \hline
1                       & 31.32 & 31.29 & \bf34.19 & 31.49 & 31.17 & 31.14 & \bf33.00    & 31.30   \\
\hdashline
2                       & 27.40  & 27.35 & \bf30.68 & 27.57 & 27.36 & 27.33 & \bf29.24 & 27.53  \\
\hdashline
3                       & 26.80  & 26.75 & \bf29.79 & 26.97 & 26.73 & 26.69 & \bf28.36 & 26.90   \\
\hdashline
4                       & 24.89 & 24.85 & \bf28.19 & 25.06 & 24.89 & 24.85 & \bf26.68 & 25.06  \\
\hdashline
5                       & 30.20  & 30.17 & \bf32.95 & 30.35 & 30.14 & 30.11 & 30.25 & \bf30.27  \\
\hdashline
6                       & 30.97 & 30.97 & \bf32.52 & 31.14 & 30.75 & 30.73 & \bf32.19 & 30.91  \\
\hdashline
7                       & 25.15 & 25.09 & \bf28.73 & 25.34 & 25.07 & 25.02 & \bf27.07 & 25.24  \\
\hdashline
8                       & 30.13 & 30.08 & \bf33.70  & 30.32 & 29.87 & 29.84 & \bf32.25 & 30.05  \\
\hdashline
9                       & 28.36 & 28.31 & \bf31.53 & 28.53 & 28.25 & 28.20  & \bf30.03 & 28.40   \\
\hdashline
10                      & 31.59 & 31.54 & \bf33.36 & 31.75 & 31.29 & 31.25 & \bf32.92 & 31.42  \\
\hdashline
average                 & 28.69 & 28.64 & \bf31.56 & 28.85 & 28.55 & 28.52 & \bf30.19 & 28.71  \\
 \hline
 \multirow{2}{*}{Image} &  \multicolumn{4} {|c|} { $\delta {\rm{ = }}25$ }&   \multicolumn{4} {|c|} { $\delta {\rm{ = }}30$}   \\    
    \cline{2-9}     
                       & CSC   & SCSC  & BM3D  & Ours  & CSC   & SCSC  & BM3D  & Ours   \\
     \hline
1                       & 30.47 & 30.46 & \bf31.97 & 30.48 & 29.20  & 29.02 & 29.18 & \bf29.29  \\
\hdashline
2                       & 27.17 & 27.16 & \bf28.34 & 27.29 & 26.42 & 26.43 & 26.44 & \bf26.46  \\
\hdashline
3                       & 26.62 & 26.59 & 26.70 & \bf26.72 & 25.97 & 26.00    & 25.98 & \bf26.00     \\
\hdashline
4                       & 24.81 & 24.78 & 24.83 & \bf24.94 & 24.42 & 24.43 & 24.45 & \bf24.49  \\
\hdashline
5                       & 29.73 & 29.73 & 29.74 & \bf29.75 & 28.55 & 28.41 & 28.59 & \bf28.63  \\
\hdashline
6                       & 30.17 & 30.18 & \bf30.69 & 30.19 & 29.00    & 29.09 & \bf29.15 & 28.84  \\
\hdashline
7                       & 24.91 & 24.89 & \bf25.23 & 25.05 & 24.49 & 24.51 & 24.53 & \bf24.57  \\
\hdashline
8                       & 29.48 & 29.47 & 29.52 & \bf29.53 & 28.32 & 28.21 & 28.34 & \bf28.39  \\
\hdashline
9                       & 27.90  & 27.89 & 27.95 & \bf27.98 & 27.18 & 27.14 & 27.20  & \bf27.23  \\
\hdashline
10                      & 30.64 & 30.62 & 30.54 & \bf30.65 & 29.30  & 29.10  & 29.36 & \bf29.38  \\
\hdashline
average                 & 28.19 & 28.18 & \bf28.55 & 28.26 & 27.29 & 27.23 & 27.32 & \bf27.33 \\
\hline
\end{tabular}
\label{table1}
\end{table*}

\begin{figure*}[htbp]
\setlength{\abovecaptionskip}{0cm}

 \begin{center}

  \subfigure[Clean image]{\includegraphics[width=0.3\textwidth]{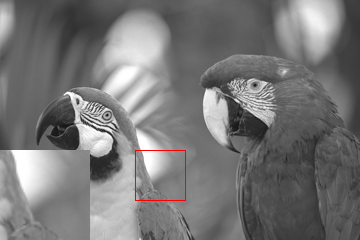}}
  \subfigure[Noisy image ($\delta {\rm{ = }}30$)]
 {\includegraphics[width=0.3\textwidth]{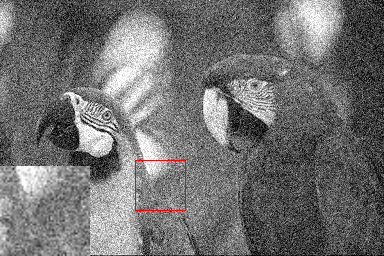}}
  \subfigure[CSC (PSNR:29.30)]{\includegraphics[width=0.3\textwidth]{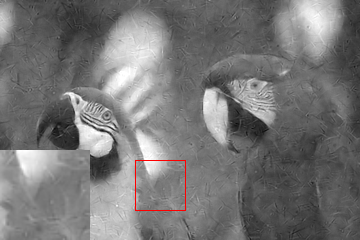}}
  \subfigure[SCSC (PSNR:29.10)]{\includegraphics[width=0.3\textwidth]{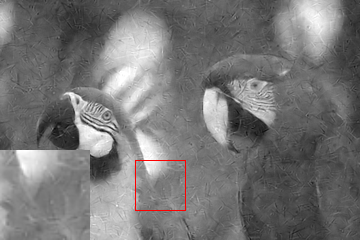}}
  \subfigure[BM3D (PSNR:29.36)]{\includegraphics[width=0.3\textwidth]{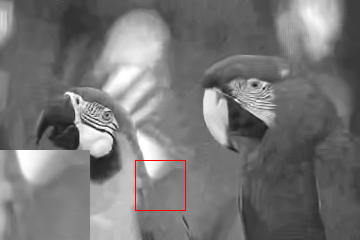}}
  \subfigure[Ours (PSNR:\bf29.38)]{\includegraphics[width=0.3\textwidth]{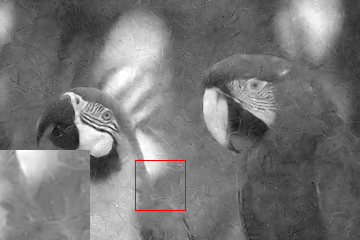}}

 \end{center}

\caption{Denoising results on image parrot from the “kodim” dataset by different methods $\delta {\rm{ = }}30$.}

\label{fig2}

\end{figure*}

\begin{figure*}[htbp]
\setlength{\abovecaptionskip}{0cm}

 \begin{center}

  \subfigure[Clean image]{\includegraphics[width=0.15\textwidth]{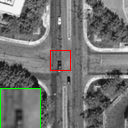}}\hspace{0.1mm}
  \subfigure[Noisy image]{\includegraphics[width=0.15\textwidth]{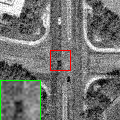}}\hspace{0.1mm}
  \subfigure[CSC(25.56 dB)]{\includegraphics[width=0.15\textwidth]{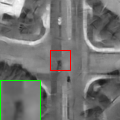}}\hspace{0.1mm}
  \subfigure[SCSC(25.51 dB)]{\includegraphics[width=0.15\textwidth]{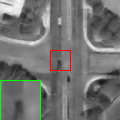}}\hspace{0.1mm}
  \subfigure[GCSC(25.40 dB)]{\includegraphics[width=0.15\textwidth]{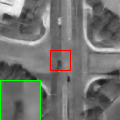}}\hspace{0.1mm}
  \subfigure[Ours(\bf25.66 dB)]{\includegraphics[width=0.15\textwidth]{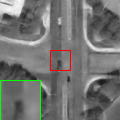}}\hspace{0.1mm}
   \subfigure[Clean image]{\includegraphics[width=0.15\textwidth]{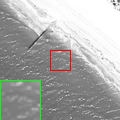}}\hspace{0.1mm}
  \subfigure[Noisy image]{\includegraphics[width=0.15\textwidth]{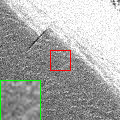}}\hspace{0.1mm}
  \subfigure[CSC(27.84 dB)]{\includegraphics[width=0.15\textwidth]{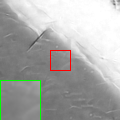}}\hspace{0.1mm}
  \subfigure[SCSC(27.81 dB)]{\includegraphics[width=0.15\textwidth]{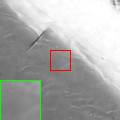}}\hspace{0.1mm}
  \subfigure[GCSC(27.74 dB)]{\includegraphics[width=0.15\textwidth]{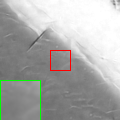}}\hspace{0.1mm}
  \subfigure[Ours(\bf27.91 dB)]{\includegraphics[width=0.15\textwidth]{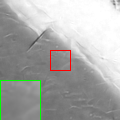}}\hspace{0.1mm}

 \end{center}

\caption{Denoising results on images "highway" and "beach" from the UVA dataset by different methods $\delta {\rm{ = }}20$.}

\label{fig3}

\end{figure*}

Table \ref{table1} shows the PSNR value comparison of the denoising results of 10 images in the "kodim" dataset with different models. The most outstanding results (highest PSNR value) at each noise level are highlighted in bold. It can be observed that the denoising performance of the model proposed in this article is comparable to that of the BM3D method. When the noise level was lower than 25$\%$, the denoising performance of BM3D was relatively high; however, when the noise level was higher than 25$\%$, the performance of the DCSC approximated or exceeded that of BM3D. This is because the BM3D algorithm mostly led to inaccurate block matching for images with high complexity or different noise intensities. Fig.\ref{fig1} shows the average denoising results of DCSC and SCSC under different noise levels. It can be observed that the PSNR value of the denoising result of the proposed model is higher than that of the SCSC. Fig.\ref{fig2} provides a visual display of the denoising results of a 768×512 image with 30$\%$ Gaussian noise in the "kodim" dataset with different denoising models. Based on the visualization results, it can be observed that the model proposed in this article achieved better denoising results than SCSC, and contained fewer noise points. Although the BM3D result appears clearer and smoother than that of the proposed model, it is excessively smooth, resulting in the loss of image details.

To further verify the performance of the model, a substantial number of images in the "drone" dataset were randomly selected for experiments, and the experimental results were compared with those of the recently proposed GCSC model that adds a gradient regular term [30] to the CSC model. Table \ref{table2} shows the average PSNR value of the denoising results of 10 images under different models, and the noise level was between 15$\%$ to 35$\%$. The best result (the highest PSNR value) under each noise level presented the most coarseness. It can be seen from Table \ref{table2} that the model proposed in this article maintained the highest PSNR value under different noise levels. Finally, a visual comparison of denoising results of "highway" and "beach" images captured by a UAV under different models is shown in Fig.\ref{fig3}, and the noise level is 20$\%$. The noise points of the model proposed in this article were less than those of the other models; therefore, the clarity of the image was improved.

\begin{table}
\setlength{\abovecaptionskip}{0cm}
\setlength{\belowcaptionskip}{0.1cm}
\renewcommand\arraystretch{1.5}
\caption{Display of the average PSNR value of the de-noising results of 10 UAV images in different models}
\label{table2}
\centering
\begin{tabular}{|m{1.2cm}<{\centering}|m{1.2cm}<{\centering}|m{1.2cm}<{\centering}| m{1.2cm}<{\centering}|m{1.2cm}<{\centering}|}  

\hline
 $\delta $ & CSC   & SCSC  & GCSC  & Ours   \\ 
\hline
15 & 26.74 & 26.57 & 26.54 & \bf26.85  \\ 
\hdashline
20 & 26.58 & 26.54 & 26.44 & \bf26.70  \\ 
\hdashline
25 & 26.36 & 26.34 & 26.26 & \bf26.43  \\ 
\hdashline
30 & 25.75 & 25.77 & 25.79 & \bf25.79  \\ 
\hdashline
35 & 24.64 & 24.75 & \bf24.92 & 24.76  \\
\hline
\end{tabular}
\label{table2}
\end{table}

The maximum number of iterations of the experiment in this article is 50. In order to make the experimental results more convincing, this article shows in Table \ref{table3} the average PSNR value corresponding to the denoising results of the 10 UAV images in Table \ref{table2} under different models when the experimental iterations are 25 times. It can be seen from Table \ref{table3} that when the experiment is iterated for 25 times, the changing trend of the denoising results of each model under different noise levels is consistent with that of Table \ref{table2}, and the model proposed in this paper still maintains the highest PSNR value.

\begin{table}
\setlength{\abovecaptionskip}{0cm}
\setlength{\belowcaptionskip}{0.1cm}
\renewcommand\arraystretch{1.5}
\caption{The average PSNR value of the denoising results of 10 UAV images in 25 iterations of different models}
\label{table3}
\centering
\begin{tabular}{|m{1.2cm}<{\centering}|m{1.2cm}<{\centering}|m{1.2cm}<{\centering}| m{1.2cm}<{\centering}|m{1.2cm}<{\centering}|}  

\hline
 $\delta $ & CSC   & SCSC  & GCSC  & Ours   \\ 
\hline
15 & 26.72 & 26.67 & 26.52 & \bf26.83  \\ 
\hdashline
20 & 26.60 & 26.54 & 26.43 & \bf26.68  \\ 
\hdashline
25 & 26.35 & 26.32 & 26.25 & \bf26.42  \\ 
\hdashline
30 & 25.64 & 25.67 & 25.70 & \bf25.71  \\ 
\hdashline
35 & 24.58 & 24.58 & \bf24.86 & 24.67  \\
\hline
\end{tabular}
\label{table3}
\end{table}

\section{Summary}
In this paper, the CSC model was integrated with the dual graph Laplacian regularizer to effectively utilize the similarity between image rows and columns. Compared with a single graph, the dual graph has a simple structure, relaxes the conditions, and is more conducive to image restoration using image signal priors. The corresponding optimization problem was formulated, and the ADMM algorithm was employed in the DFT domain. Finally, denoising experiments were performed using Gaussian white noise. The results proved that the model proposed in this article is superior to other models integrated with a single graph Laplacian regularization term. However, this article also has shortcomings, which are also areas that need to be improved in the future. Although the introduction of the dual graph improves the denoising performance of the original model to a certain extent, the dual graph is constructed from the original input image. Its structure is not updated with the update of the feature map, so there is no guarantee that each feature map has the structure of the map defined on the input image. From a theoretical analysis, if the dual graph is constructed on the feature graph, then it is constantly updated according to the iteration of the algorithm, which can further limit the smoothness of the feature graph on the graph and make the model obtain better performance. This perspective can be used as a research direction of the CSC model in the future. 

%% The Appendices part is started with the command \appendix;
%% appendix sections are then done as normal sections
%% \appendix

%% \section{}
%% \label{}

%% If you have bibdatabase file and want bibtex to generate the
%% bibitems, please use
%%
%%  \bibliographystyle{elsarticle-num} 
%%  \bibliography{<your bibdatabase>}

%% else use the following coding to input the bibitems directly in the
%% TeX file.

\end{document}